\documentclass[%
 reprint,
 superscriptaddress,
 amsmath,amssymb,
 aps,
]{revtex4-2}

\usepackage{graphicx}  
\usepackage{dcolumn}   
\usepackage[svgnames, table]{xcolor}    

\usepackage{amsfonts}
\usepackage{comment}
\usepackage{tabularx}
\usepackage{enumitem}
\usepackage{framed}
\usepackage[colorlinks=true,linkcolor=blue, urlcolor=black, citecolor=blue]{hyperref}
\usepackage{ragged2e}

\definecolor{shadecolor}{gray}{0.97}

\newcolumntype{Y}{>{\centering\arraybackslash}X}

\begin{document}

\preprint{APS/123-QED}

\title{Minimal motifs for habituating systems}

\author{Matthew Smart}
\email{msmart@flatironinstitute.org}
\affiliation{Center for Computational Biology, Flatiron Institute, New York, NY}

\author{Stanislav Y. Shvartsman}
\email{stas@princeton.edu}
\affiliation{Center for Computational Biology, Flatiron Institute, New York, NY}
\affiliation{Department of Molecular Biology, Princeton University, Princeton, NJ}
\affiliation{Lewis-Sigler Institute for Integrative Genomics, Princeton University, Princeton, NJ}

\author{Martin M\"onnigmann}
\email{martin.moennigmann@rub.de}
\affiliation{Department of Mechanical Engineering, Ruhr-Universit\"at Bochum}



\begin{abstract}
    Habituation – a phenomenon in which a dynamical system exhibits a diminishing response to repeated stimulations that eventually recovers when the stimulus is withheld – is universally observed in living systems from animals to unicellular organisms. 
    Despite its prevalence, generic mechanisms for this fundamental form of learning remain poorly defined. 
    Drawing inspiration from prior work on systems that respond adaptively to step inputs, we study habituation from a nonlinear dynamics perspective. 
    This approach enables us to formalize classical hallmarks of habituation that have been experimentally identified in diverse organisms and stimulus scenarios. 
    We use this framework to investigate distinct dynamical circuits capable of habituation. 
    In particular, we show that driven linear dynamics of a memory variable with static nonlinearities acting at the input and output can implement numerous hallmarks in a mathematically interpretable manner.
    This work establishes a foundation for understanding the dynamical substrates of this primitive learning behavior and offers a blueprint for the identification of habituating circuits in biological systems.
\end{abstract}

\maketitle

To survive and reproduce, living systems must sense and respond to changes in their environment.
Benign stimuli that are repeatedly presented must be recognized and ignored in order to conserve energy and focus on more important tasks. 
The capacity to detect and ``tune out" distractions, as quantified by a progressively diminishing response to subsequent stimulation, is known as habituation and is universally observed across living systems \cite{TANG2018R1180}. 
To illustrate the phenomenon, consider the response of an organism to repeated dimming (dark-flashes) mimicking a sudden threat. 
When subjected to such stimuli, fruit flies initially exhibit a jump response before eventually becoming passive \cite{Engel2009}. 
Similarly, larval zebrafish will learn to ignore repeated dimming despite rapidly reorienting their motion in response to the initial stimulation \cite{Randlett2019}. 
When organisms that are habituated to a particular stimulus are presented with a sufficiently different one, the original response is immediately recovered (termed \emph{dis}habituation) \cite{Thompson1966, rankin2009}; this rules out trivial mechanisms such as fatigue and indicates that neutral stimuli are indeed detected and ignored. 

Habituation is recognized as a fundamental form of nonassociative learning \cite{TANG2018R1180}. 
Is is commonly deregulated in neurodevelopmental disorders \cite{Blok2022}, and 
has long been studied from a neurological perspective \cite{Thompson2009}. 
Early experimental works characterizing this phenomenon focused on the neural substrates of habituating responses in animals such as the sea slug \emph{Aplysia} \cite{Carew1972, Carew1979} (see \cite{Carew1986annualrev} for a review focusing on invertebrates).
The hallmarks associated with habituation in living systems were cataloged about fifty years ago \cite{Thompson1966} and have since been revisited and refined \cite{rankin2009}.
Beyond habituation itself, a key hallmark is spontaneous forgetting (recovery), whereby a system's capacity to respond eventually recovers when the stimulus is withheld for a sufficiently long time. 
Other notable hallmarks consider how the rapidity of habituation depends on the intensity or the frequency of stimuli, along with dishabituation (mentioned above).
A central aim of this work is the identification of minimal dynamical systems which recapitulate specific hallmarks.

The capacity for habituation and other forms of learning is not exclusive to multi-neuron systems \cite{TANG2018R1180, Gunawardena2022}. 
Supporting this, there is ample evidence that individual neurons can habituate \emph{in vitro} \cite{Mcfadden1990, Cheever1994}.
Remarkably, even unicellular organisms -- lacking conventional neural components -- can learn from repeated stimuli and exhibit habituation. 
This insight dates back to the work of Jennings at the turn of the twentieth century \cite{Jennings1906} and is enjoying renewed interest \cite{Gershman2021}, due in part to modern experimentation able to address reproducibility concerns and inter-individual variability. 
For instance, recent experiments with the giant ciliate \emph{Stentor coeruleus} \cite{Rajan2023} demonstrate habituation of a contractile response to mechanical vibration. 
Additional evidence of non-neuronal habituation has been reported in other ciliates \cite{Patterson1973, Eisenstein1982}, in the slime mold \emph{Physarum polycephalum} \cite{Boisseau2016, Boussard2019}, as well as the plant \emph{Mimosa pudica} \cite{Gagliano2014}.
These findings in organisms without neurons (let alone what one might call a brain) indicate that primitive forms of learning can be implemented by other means, perhaps biochemically. 
Pushing this direction even further, recent studies of the conductive properties of certain materials \cite{Zuo2017, Zhang2021resistance} subjected to repeated hydrogen exposure demonstrate that non-living systems can also exhibit habituation and recovery.
Taken together, these works point to the generality of the habituation phenomenon and expand the scope of a key open question: what is the minimal description of how memory is dynamically established, maintained, and read-out across such diverse settings?

\begin{figure*}[ht]
    \centering\includegraphics[width=0.95\linewidth]{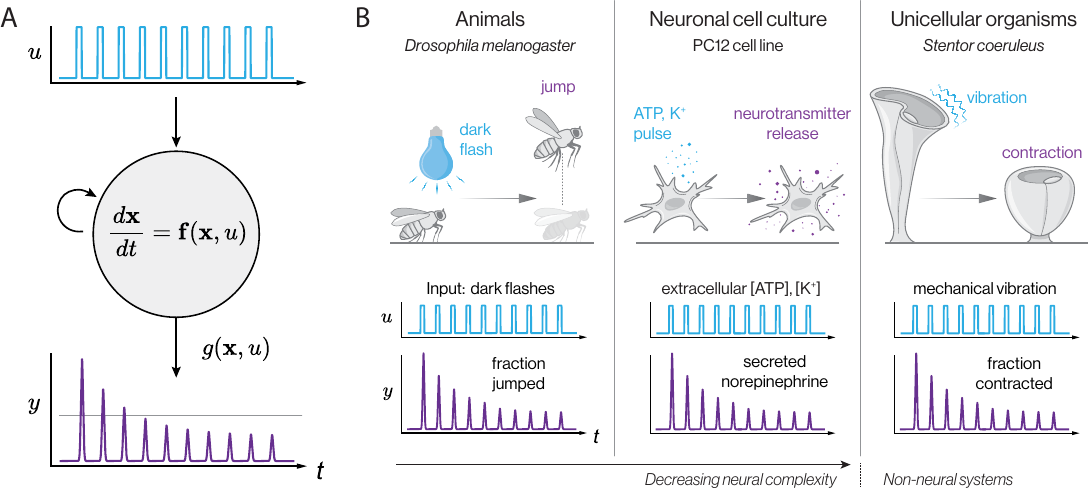}
    \caption{
    (A) Habituating system input-output behavior. 
    Application of a pulsatile, periodic stimulus $u(t)$ to the nonlinear system 
    $d\mathbf{x}/dt=\mathbf{f}(\mathbf{x},u)$ results in a habituating response $y(t)=g(\mathbf{x},u)$ characterized by monotonically decreasing output peaks 
    (see Hallmark 1 of Table \ref{tab:hallmarks-main}). 
    (B) Habituation is observed across diverse biological systems and scales spanning from animals with complex nervous systems (e.g. escape response of fruit flies \cite{Engel2009}, zebrafish \cite{Randlett2019}), 
    to individual neurons in culture 
    (e.g. rat pheochromocytoma cells \cite{Mcfadden1990, Cheever1994}), 
    to unicellular organisms (e.g. the freshwater ciliate \emph{Stentor coeruleus} \cite{Rajan2023}).
    }
    \label{fig:fig1_outline}
\end{figure*}

Models for habituation have been historically motivated by neural implementations in particular experimental contexts 
\cite{Stanley1976, Staddon1993, Staddon1996, Deliang1993, Dragoi2002, Sirois2004, Schoner2006}, 
leading to many system-specific details and assumptions 
(see e.g. the review \cite{Sirois2002}).
Acknowledging that habituation occurs in organisms lacking nervous systems, we focus here on generic dynamical systems that could conceivably be implemented in single cells. 
The literature in this area is comparatively limited 
\cite{Bonzanni2019, eckert2022}. 
Notably, the recent thesis of Eckert \cite{eckert2022} studied the phenomenon from a biochemical perspective,
formalized several of the hallmarks so that they could be numerically screened, and proposed variants of an incoherent feedforward motif that satisfied one or more hallmarks. That work, in turn, was motivated by two works of Staddon and Higa in the 1990s \cite{Staddon1993, Staddon1996}, which show that a simple discrete-time integrator with nonlinear output can exhibit habituation and recovery, and that chaining these units in a feedforward manner can additionally exhibit frequency sensitivity. Our aims are similar to these works in that we seek to find the simplest possible dynamical systems that satisfy various hallmarks of habituation. 
Finally, we note that there is an extensive body of work studying biochemical implementations of the closely related phenomenon of adaptation, 
which we draw inspiration from and address later.

Taken together, the observations above beg the question of what generic, biologically implementable circuits could be responsible for mediating habituation across biological scales and functional contexts. 
Specifically, we aim to identify simple single-input, single-output motifs whose deterministic dynamics exhibit habituation and the associated hallmarks.
Our contributions are as follows: First, we provide a mathematical characterization of the classic habituation hallmarks (see Table \ref{tab:hallmarks-main}). In search of systems that satisfy these hallmarks, we draw a connection between habituation and adaptation and characterize the habituation potential of known adaptive systems. We then consider the simplest possible systems, noting that linear systems do not habituate. We identify a minimal motif involving linear memory dynamics with static output nonlinearities that satisfies the basic hallmarks and is readily implementable. We then show how simple extensions of the proposed system can cover additional hallmarks, and close with a discussion and suggestions for future work.

\section{Results}

\subsection{Mathematical formulation of the hallmarks}
Ten hallmarks of habituation observed across animals (focusing on neurobehavioral aspects) are described in \cite{Thompson1966, rankin2009}. 
We list them in shortened form in Table~\ref{tab:hallmarks-main}.
In addition to the core hallmarks of habituation itself and spontaneous recovery, these include more detailed features like frequency and amplitude sensitivity. 

Our first goal is to establish a framework from which the hallmarks can be stated in mathematical terms, enabling the identification of minimal dynamical models in subsequent sections.
As we mainly restrict to a single-input, single-output setting, we do not focus on aspects involving multiple inputs (specifically 7, 8, 9). 
Using the notation introduced below, we propose mathematical criteria alongside the classical hallmarks in Table \ref{tab:hallmarks-main}. 

\begin{figure*}[ht]
    \centering\includegraphics[width=0.99\linewidth]{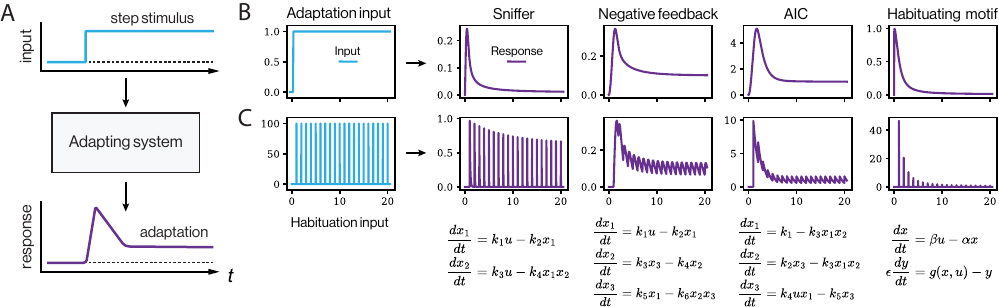}
    \caption{
    Systems capable of adaptation to step inputs can also exhibit habituation to pulsatile inputs, and vice-versa. 
    (A) The asymptotic response of an adapting system to a step input returns towards its original equilibrium.
    (B) Known adapting motifs include incoherent feedforward loops such as the Sniffer  \cite{Tyson2003-SniffersBuzzersEtc}, negative feedback loops \cite{Ma2009}, and antithetical integral control (AIC) \cite{Ferrell201662}. 
    The minimal habituating motif Eq. (\ref{eq:filter_A_ode}) with $g(x,u)=u/(1+x^2)$ (see Section \ref{sec:construct_filter_motif}) also adapts to step inputs. 
    (C) Each system exhibits habituation to pulsatile inputs (duty $d=10^{-2}$) for the same system parameters.
    The final dynamical variable is used as the response (output) for each system. 
    Parameters are set to $1$ except: 
    (Sniffer) $k_2=0.1, k_4=10$; 
    (NF) $k_4=0.1, k_6=10$;  
    (AIC) $k_4=10$; and 
    $\alpha = 0.1, \epsilon=10^{-2}$.
    }
    \label{fig:adapt_vs_hab}
\end{figure*}

\newpage
In the spirit of motifs~\cite{Alon2006, Goentoro2009}, we consider models of the form $\frac{d \mathbf{x}}{dt}= \mathbf{f}(\mathbf{x}, u)$, where $\mathbf{x} \in\mathbb{R}^n$ denotes internal state variables and $u(t)$ is a scalar input to the system. 
We model stimuli by non-negative periodic rectangular pulses with period $T$, amplitude $A$, and duty $d \in (0,1)$. 
The pulse area $\Lambda = \int_t^{t+T} u(\tau) d\tau$ serves as a measure of the intensity of the input signal. 
The system produces a scalar \emph{response} $y(t)= g(\mathbf{x}, u)$. 
Let $\theta\in\mathbb{R}^p$ denote the parameters of $\mathbf{f}$, $g$ such as rate constants.
Our aim is to identify simple choices of $\mathbf{f}$ and $g$ such that $y(t)$ satisfies as many hallmarks as possible, in a manner that is structurally stable with respect to $\theta$ and stimulus properties. 
Since all biological quantities are bounded, we may assume without restriction they are bounded from below by zero. 
Recovery is studied with signals $u(t)$ that alternate between sequences of pulses and waiting periods with $u(t)= 0$. 
We assume there exists a steady state with response zero ($y=0$) in the absence of a stimulus ($u=0$). 
Because periodic signals frequently appear, it is convenient to denote time in multiples of the period of interest $T$. 
We write $y[k]$, $k= 0, 1, \dots$ as shorthand notation for the highest value in period $k$
($y[k]= \max_{[kT, (k+1)T)} y(t)$). 

With the framework established above, we can now search for candidate dynamical systems that exhibit specific hallmarks (defined mathematically in Table \ref{tab:hallmarks-main}). We start with the first hallmark – often called habituation itself – which roughly states that pulsatile inputs $u$ generate responses $y$ with decreasing peaks. We first show that various motifs for adaptation respect $H_1$ and thus may serve as a guide to constructing a minimal motif for habituation. 

\subsection{Motifs for adaptation can exhibit habituation}

In addition to habituation, a ubiquitous feature of biological systems is adaptation, or homeostasis, which can be defined as the ability to maintain a specific target state in response to a slowly shifting environment. 
For instance, an organism might need to maintain an internal body temperature $y^*$ for a finite range of environmental temperatures $u(t)$. 
As pointed out in \cite{eckert2022} and suggested by Fig. \ref{fig:fig1_outline}A and Fig. \ref{fig:adapt_vs_hab}A, adaptation and habituation are naturally related concepts; this relationship is 
subtle, however, because adaptation is a feature of the asymptotic ($t\rightarrow\infty$) response whereas habituation concerns both the transient and asymptotic behavior.

In our formalism, we say that the system $d\mathbf{x}/dt=\mathbf{f}(\mathbf{x},u), \quad y=g(\mathbf{x},u)$ exhibits \emph{perfect adaptation} if $|g(\mathbf{x},0) - g(\mathbf{x},u)| \rightarrow 0$ as $t \rightarrow \infty$. This means that the output $y(t)$ associated with a non-zero input $u(t)$ asymptotically returns to the zero-input value $y^*$. 
This is non-trivial because it does not assume $u(t)\rightarrow 0$ asymptotically, 
with step inputs ($u(t) = A H(t)$, scaled Heaviside step function) being the primary class considered in prior literature. 
The identification and characterization of molecular circuits that can implement adaptation in this general setting have been the subject of significant study \cite{Tyson2003-SniffersBuzzersEtc, Ma2009, Ferrell201662, Briat201615, Khammash2019}; see also alternative formulations from Golubitsky et al. 
\cite{Golubitsky2017, Reed2017, Golubitsky2020-InfHomeo3Nodes}. 
Due to the close connection between habituation and adaptation, many adapting systems can also habituate even when the system parameters $\theta$ are fixed.
Several examples are provided in Fig. \ref{fig:adapt_vs_hab}B and we return to this point in the Discussion.

While these examples of adapting systems establish candidate models of habituation, they do not address the question of what the \emph{simplest} model for such behavior is. 

\subsection{Linear time-invariant systems do not habituate}
Choosing $\mathbf{f}$ and $g$ to be linear in $\mathbf{x}$ and $u$ is arguably the simplest choice one could make. 
It can be shown, however, that linear systems are not able to habituate. This result holds for a very general class of linear time-invariant systems \cite{Monnigmann2024CDC}, which includes but is not restricted to the finite-dimensional ordinary differential equations treated here and in the literature on motifs~\cite{Alon2006, Tyson2003-SniffersBuzzersEtc,Ferrell201662}. It is evident that linear systems are not appropriate, because responses of biological systems are bounded (consider e.g. non-negative concentrations) while linear systems respect the superposition principle and thus are unbounded. Even if bounded responses can be guaranteed for a linear system for bounded stimuli (cf. e.g. the concept of bounded input bounded output stability), the superposition principle of linear time-invariant systems contradicts hallmark $H_1$. 
We depict this argument graphically in Fig. \ref{fig:linear_fails_and_minimal_model} and refer the reader to \cite{Monnigmann2024CDC} for details. 
Linear systems therefore do not habituate, raising the question of what the next simplest structure for habituation could be. 

\subsection{Constructing a simple system that displays multiple hallmarks of habituation}
\label{sec:construct_filter_motif}

\begin{figure*}[ht]
    \centering\includegraphics[width=0.999\linewidth]{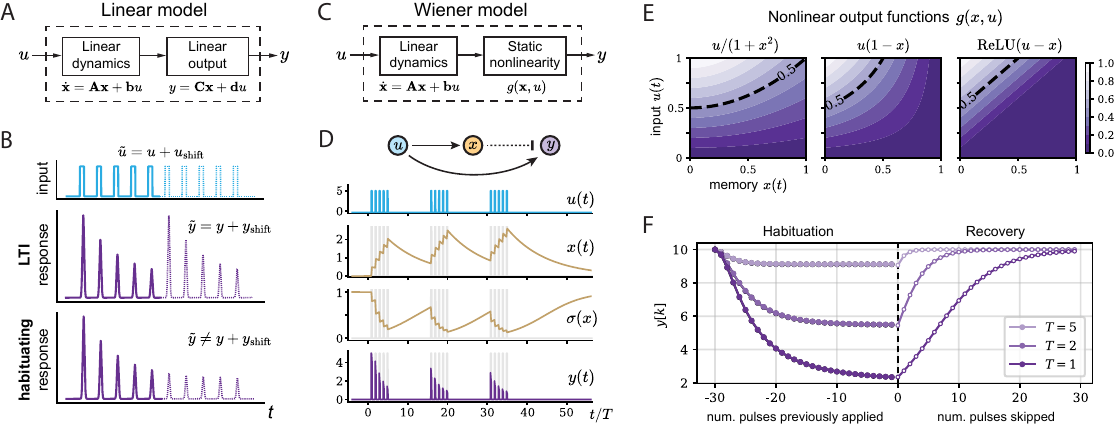}
    \caption{
    (A, B) Linear and Wiener systems both assume linear state space dynamics $\mathbf{f}(\mathbf{x},u)=\mathbf{Ax}+\mathbf{b}u$. The former restricts to linear outputs, while the latter allows any $g(\mathbf{x},u): \mathbb{R}^{n+1} \rightarrow \mathbb{R}$. 
    (C) Linear time-invariant systems do not respect Hallmark 1, as depicted using a linear superposition of hypothetical responses to shifted inputs which habituate when considered separately. 
    (D) The model Eq. (\ref{eq:filter_A_ode}) can be represented graphically as an incoherent feedforward motif. 
    Example behavior for three cycles of five rectangle pulses ($T=1, d=0.2, A=5$) followed by ten rest periods ($L= 5$, $L^\prime= 10$). 
    The static nonlinearity is $g(x,u)=u \sigma(x)$ with $\sigma(x)=1/(1+x^2)$ and parameters $\alpha=0.1, \beta=0.5$. Qualitatively similar behavior is observed for gaussian pulses and other choices for $g(x,u)$.
    (E)
    Example output functions $y=g(x,u)$ which attenuate an
    input signal $u(t)$ using a memory state $x(t)$. 
    Each functional form satisfies 
    (i) Inputs are not suppressed ($y \approx u$) when $x$ is low; 
    (ii) The input signal is attenuated (weak or no system response) when $x$ is high.
    (F) System exhibits partial frequency sensitivity, Hallmark 4(a) (Table \ref{tab:hallmarks-main}). 
    Solid circles denote the responses to $30$ consecutive pulses applied at different periods $T$.
    Open circles denote the response to a pulse after waiting $k$ periods (omitting $k$ pulses before applying another).
    The vertical dashed line shows the point from which recovery is measured. 
    The curves are given by $y[k]= A \sigma(x[k])$.  
    Left (habituation): $x[k] = Q \frac{1-q^{k}}{1-q} q^{1-d}$ is the value of $x$ immediately before the $k^{th}$ pulse, with $q \equiv e^{-\alpha T}$ and $Q\equiv A \frac{\beta}{\alpha}(1-q^d)$ (see Materials and Methods).
    Right (recovery): $x[k] = x_0 q^k$ where $x_0$ is the value after $30$ pulses.
    System: $\alpha=0.1, \beta=0.2$. 
    Stimuli: amplitude $A=10$ and area $A d T=1$ are fixed while $T$ varies. 
    }
    \label{fig:linear_fails_and_minimal_model}
\end{figure*}

Towards identifying simple systems that habituate and recover, we start by considering output functions of the form $y(t)= u(t) \sigma(t)$.
The \emph{receptivity} $\sigma(t)$ acts as a multiplicative filter on the input $u(t)$. 
In this setting, our goal is to find such filters that turn periodic signals $u(t)$ into signals $y(t)$ that satisfy Hallmarks 1 and 2. 

An expedient approach is to scale the input by an exponential envelope $\sigma(t) = e^{-\alpha t} H(t)$, where $H$ denotes the Heaviside step function. 
For a periodic signal starting at $t=0$, $y(t)= u(t) \sigma(t)$ will trivially satisfy Hallmark 1 (habituation). 
This is not biologically sensible, however, because there is no reason to privilege one specific time ($t=0$) over another. 
Furthermore, this will not satisfy Hallmark 2 (recovery) because stimuli at long times $t\gg 1/\alpha$ are suppressed irrespective of the recent past. 

How might short-term memory (forgetting) be incorporated in $\sigma(t)$? 
A more biologically realistic extension of the intuition above achieves this. 
First, introduce an internal system state that convolves the input with a truncated exponential kernel: 
\begin{equation}
  \label{eq:filter_A_convolve}
  x(t) = u(t) * \beta e^{-\alpha t}H(t) = \int_{-\infty}^t \beta e^{-\alpha (t-\tau)} u(\tau) d\tau
\end{equation}
From this, define a sigmoidal attenuation function such as $\sigma(x(t))=1/(1+x^N)$ for fixed $N\ge 1$. 
The idea is two-fold:
First, $x(t)$ performs a weighted sum of recent inputs such that older inputs are forgotten with rate $\alpha$. 
Second, the receptiveness of the system to imminent inputs, $\sigma(x)$, is determined by biologically plausible saturation of $x$ (e.g. via phosphorylation \cite{Gunawardena2005}), and satisfies $\sigma(0)=1$ with monotonic decrement as $x\rightarrow \infty$. 
In this way, the response $y(t)=u(t)\sigma(x)$ ignores new stimuli when the system is saturated, while fully responding to new stimuli when few recent ones have been observed.
See Fig. \ref{fig:linear_fails_and_minimal_model}D for an example demonstrating habituation and recovery.

Differentiating Eq. (\ref{eq:filter_A_convolve}) reveals a linear ODE for the memory variable:
$dx/dt = \beta u - \alpha x$.
We thus have linear dynamics with a static nonlinear output, 
placing it in the class of Wiener models \cite{Schoukens2017} 
(Fig. \ref{fig:linear_fails_and_minimal_model}C).
Putting this together, a minimal model for habituation and recovery is the one-dimensional system,
\begin{equation}
\label{eq:filter_A_ode}
\begin{array} {rcl} 
             dx/dt & = & \beta u - \alpha x\\
               y(t) & = & g(x,u), \\
\end{array}
\end{equation}
where $g(x,u)=u/(1+x^N)$ as defined above. 
Given an arbitrary stimulus $u(t)$ and $x(t_0)=x_0$, we have $x(t)= x_0e^{-\alpha \left(t-t_0\right)}+ \beta\int_{t_0}^te^{-\alpha(t-\tau)} u(\tau)d \tau$ (with Eq. (\ref{eq:filter_A_convolve}) recovered when $x(-\infty)=0$).

This motif is structurally robust in that the particular form of the output, the multiplicative filter, is not critical. 
Essentially, the output should pass the input to the output when the memory variable $x$ is low and attenuate it when $x$ is high;
Fig. \ref{fig:linear_fails_and_minimal_model}E illustrates this and shows that other forms are possible. 
As an alternative to the multiplicative filter, the memory variable $x$ may be used to lower a threshold above which the input is passed to the output.
Such threshold forms are amenable to particularly simple physical implementations of habituation dynamics \cite{Monnigmann2024CDC}, which have recently attracted attention in neuromorphic computing~\cite{Zuo2017, Zhang2021resistance, Ramanathan2024}. 
Moreover, for the choice $g(x, u)= \text{ReLU}(u - x)$ (where $\textrm{ReLU}(z)=\max (z,0)$), the model maps directly onto a discrete-time model suggested by Staddon \cite{Staddon1993, Staddon1996} (see Materials and Methods). 
We return to this connection in Section \ref{sec:freq_sens}. 

Several other generalizations of our simple model are obvious. 
First, the exponential kernel in Eq. (\ref{eq:filter_A_convolve}) could be replaced with something more complex, at the potential cost of complicating the dynamics for $x(t)$. 
Second, a linear output, if desired, can be achieved by absorbing the nonlinearity into an additional nonlinear dynamic equation $\epsilon\, d x_2/dt= g(x, u)- x_2$, where $\epsilon\ll 1$ controls how quickly the now linear output $y= x_2$ approaches $g(x, u)$.

\subsubsection{System exhibits recovery and potentiation \\(Hallmarks 2 and 3)}
It is evident from Fig.~\ref{fig:linear_fails_and_minimal_model}D that the system from~\eqref{eq:filter_A_ode} recovers in periods without signal. 
The recovery is partial after the shown ten rest periods in that the response to the first new signal is weaker than the initial response (i.e. $y[0] > y[15], y[30]$).
This occurs because the receptivity $\sigma$ has not returned to $1$. 
Asymptotically however $x(t) \rightarrow 0$ and the system approaches full recovery, as shown for different stimulation periods in Fig.~\ref{fig:linear_fails_and_minimal_model}F. 

The system also displays potentiation of habituation (Hallmark 3), which essentially states the effect described with Hallmark 1 becomes stronger when stimulation and recovery happen repeatedly. This is also evident from Fig.~\ref{fig:linear_fails_and_minimal_model}, where the responses to the second series of stimuli are weaker than to the first (specifically, $K= 1$, $L= 5$, $L^\prime= 10$ in $H_3$ in Table~\ref{tab:hallmarks-main}). 
This effect can be amplified by using an output function $g(x,u)$ with a tunable response threshold (Fig. S1).

\subsubsection{\hspace{-0.08cm}Sufficient conditions for frequency sensitivity, Hallmark 4}
\label{sec:freq_sens} 

\begin{figure*}[ht]
    \centering\includegraphics[width=0.95\linewidth]{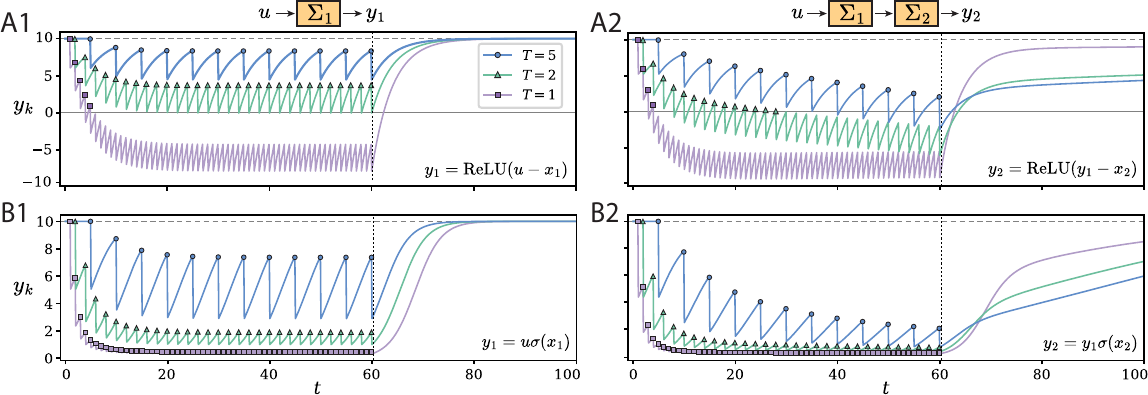}
    \caption{
    Frequency sensitivity for a single unit ($K=1$, left two panels A1, B1) and two units in series ($K=2$, right two panels A1, B2). Markers denote the response to periodic stimuli for different periods. Curves indicate receptivity, i.e. the hypothetical response to a stimulus at any time. Stimuli lead to an immediate drop in receptivity and no response occurs below the threshold (horizontal line) in A1, A2. Stimuli are applied until $t=60$ (vertical line) after which recovery is monitored. Hallmark 4(b) (more frequent stimulation results in faster recovery) does not hold for the single unit, regardless of the output function ($g(x, u)= \textrm{ReLU}(u-x)$ in A1 and $g(x, u)= u/(1+x^2)$ in B1); the receptivities all approach the same asymptote after $t=60$ but this takes longer for higher stimulus frequencies. In contrast, Hallmark 4(b) holds if two units with different time scales are connected in series, regardless of the output function (A2, B2); the receptivity curves cross, indicating that more frequent stimuli result in faster recovery. 
    Duty $d$ is chosen such that $AdT=1$ for all $T$.
    Parameters in (A): $\alpha_1=0.25, \alpha_2=5 \cdot 10^{-3}$, $\beta_1=4, \beta_2=1$.
    In (B): $\alpha_1=0.2, \alpha_2=0.02$,  $\beta_1=1, \beta_2=0.5$.  
    }
    \label{fig:figH4iii-recovery}
\end{figure*}

Others have reported that series extensions (Fig. \ref{fig:linear_fails_and_minimal_model}F) of similar models are important for implementing frequency sensitivity \cite{Staddon1996, eckert2022}, which states that more frequent stimulation results in (a) stronger habituation and (b) faster recovery (Table \ref{tab:hallmarks-main}).
Here we clarify the sufficient ingredients for both aspects in our formalism.

Regarding $H_4$(a) -- more frequent stimuli lead to stronger habituation -- we find that series connections are not necessary. 
Fig. \ref{fig:linear_fails_and_minimal_model}F and
Fig. \ref{fig:figH4iii-recovery} ($K=1$, panels A1, B1) 
show that more frequent stimuli (smaller period $T$) lead to more rapid and more pronounced habituation (``rapid" refers to the number of pulses needed to reach an arbitrary response decrement, rather than asymptotic levels as in \cite{eckert2022}).
In the proposed model the mechanism for $H_4$(a) is transparent: the memory variable $x(t)$ has less time to decay between subsequent stimulations, which results in a stronger attenuation of the input signal when passed through the static output nonlinearity.  
Thus, this aspect of frequency sensitivity can be achieved with a single unit (i.e. the system from Eq. (\ref{eq:filter_A_ode})).

The second aspect of frequency sensitivity, $H_4$(b),
was added in Rankin et al.'s revision \cite{rankin2009} to the original hallmarks \cite{Thompson1966} following experimental observations of more rapid recovery for more frequent stimulation (see e.g. \cite{Rankin1992}).
This characteristic suggests additional timescale(s) in the dynamics.
Indeed, Fig. \ref{fig:linear_fails_and_minimal_model}F and Fig. \ref{fig:figH4iii-recovery}A indicate that a single scalar unit does not exhibit $H_4$(b). 

However, in agreement with previous work, Fig. \ref{fig:figH4iii-recovery}B shows that connecting units in series is sufficient for capturing the second aspect of frequency sensitivity, faster recovery. 
We assume each unit has an output $y_k = g(x_k, u_k)$ acting which is treated as input to the subsequent unit. 
The dynamical system in the minimal case of $K=2$ units in series is
\begin{equation}
\label{eq:H4(b)}
\begin{array} {rllc} 
        d x_1 /dt & = &  \beta_1 u        -\: \alpha_1 x_1 \\
        d x_2 /dt & = &  \beta_2 g(x_1,u) - \: \alpha_2 x_2 \\
             y(t) & = &  g \left( x_2, y_1 \right). \\
\end{array}
\end{equation}
Fig. \ref{fig:figH4iii-recovery} 
demonstrates that Eq. (\ref{eq:H4(b)}) satisfies both $H_4$(a) and $H_4$(b). 
In particular, as suggested in \cite{Staddon1996}, when $\alpha_2 < \alpha_1$ 
the system acts like a nonlinear low-pass filter: frequent stimuli quickly saturate $x_1$ which prevents transmission to $x_2$, whereas less frequent stimuli can pass through to the second unit and therefore experience slower recovery. 

Overall, Fig. \ref{fig:figH4iii-recovery} shows that habituation is faster and more pronounced for more frequent stimulation in both cases ($K=1,2$), while recovery is only faster when units are chained in series ($K=2$). 
Thus, series connections are sufficient for $H_4$(b) but are not necessary for $H_4$(a), clarifying previous work which used either discrete-time models \cite{Staddon1996} or more complex dynamics \cite{eckert2022}. 

\subsubsection{Related systems which exhibit subliminal accumulation and dishabituation (Hallmarks 6 and 8)}

Two additional hallmarks can be accounted for 
using Eq. (\ref{eq:filter_A_ode}) directly or with minor modifications 
in a similar approach to \cite{Monnigmann2024CDC}.

\begin{figure}[bt]
    \centering\includegraphics[width=0.99\linewidth]{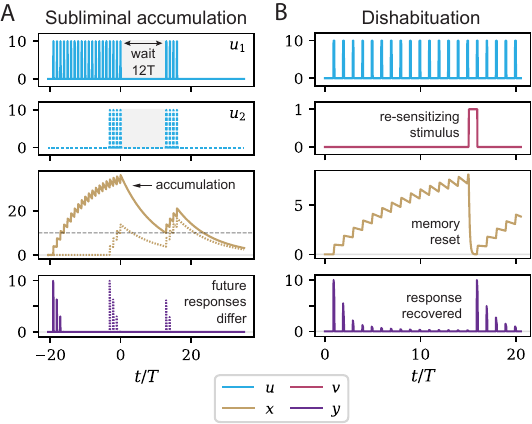}
    \caption{
    Related systems for Hallmarks 6 and 8.
    (A) Demonstration of subliminal accumulation (Hallmark 6). 
    Given the response function $g(x,u)=\textrm{ReLU}(u-x)$, the output is zero whenever $x$ exceeds $u$ (dashed black line). 
    (B) Demonstration of dishabituation (Hallmark 8). 
    In both panels: $d=0.1, T=1, A=1$, $\alpha=0.1$; $\beta=4$ in A; and $\beta=1, N=2, \kappa=10$ in B.
    For qualitatively similar behavior implemented using a more detailed physical system (electrical circuit), see \cite{Monnigmann2024CDC}.  
    }
    \label{fig:H6H8}
\end{figure}

Subliminal accumulation (Hallmark 6) states that ``the effects of repeated stimulation may continue to accumulate even after the response has reached an asymptotic level". 
This may be implemented using a response function with explicit thresholding such as $y= \textrm{ReLU}(u - x)$ introduced in Section~\ref{sec:construct_filter_motif}.
A demonstration of Hallmark 6 is shown in Fig. \ref{fig:H6H8}A. 
Observe that even after the response has reached asymptotic levels (i.e. $y=0$) the memory variable can continue to accumulate; this increases the time it takes for an overstimulated system to recover, a signature of subliminal accumulation.

Next, dishabituation (Hallmark 8) states that ``the presentation of another (usually strong) stimulus results in recovery of the habituated response".
This requires the introduction of a second stimulus, denoted $v(t)$ and treated as binary (either present $v=1$ or absent $v=0$), which can re-sensitize the system. 
We consider the modification to Eq. (\ref{eq:filter_A_ode}),
\begin{equation}
    \label{eq:hallmark5_system}
    \begin{array} {rcl} 
               d x/dt & = & \beta u - \alpha x - \kappa v x\\
                    y & = & g(x,u), \\
    \end{array}
\end{equation}
where the additional term $\kappa v x$ causes depletion of $x$ in the presence of the dishabituating stimulus $v$. 
When $\kappa$ is large, $v$ need only be presented for a short time to achieve the dishabituating effect (Fig. \ref{fig:H6H8}B). One can view the presentation of $v$ as erasing any history of recent stimulation that is stored in the memory variable $x$, thereby releasing its attenuation of the output variable $y$.

\subsubsection{System can be extended to implement amplitude sensitivity (Hallmark 5)}

\begin{figure}[t]
    \centering\includegraphics[width=0.95\linewidth]{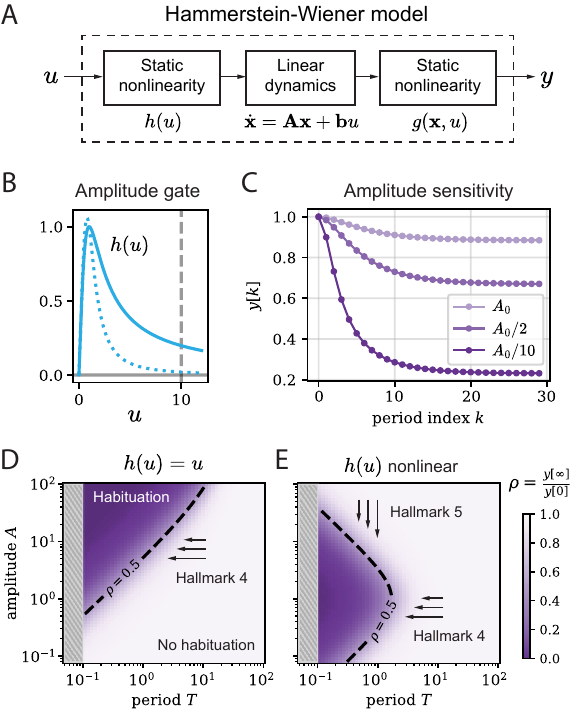}
    \caption{
    Extension of the system exhibits amplitude sensitivity.
    (A) The proposed extension Eq. (\ref{eq:hallmark5_system}) belongs to the class of Hammerstein-Wiener models. 
    (B) Amplitude gate with functional form $h(u)=2u/(1+u^N)$ for $N=2$ (solid) and $N=3$ (dashed).
    (C) When the system is driven by stimuli of different amplitudes $A$, 
    Hallmark 5 (as defined in Table \ref{tab:hallmarks-main}) is observed. 
    Stimulus parameters are $T=1, d=0.1, A_0=10$ and system parameters are $\alpha=0.2, \beta=4$, $N=2$.
    (D, E) The attenuation ratio $\rho \equiv y[\infty] / y[0]$ is shown for Eq. (\ref{eq:filter_A_ode}) in D; and for Eq. (\ref{eq:hallmark5_system}) in E. 
    Parameters follow Fig. \ref{fig:linear_fails_and_minimal_model}G and panel C. 
    When varying period $T$, duty $d$ is co-varied to fix the intensity (pulse area $\Lambda = AdT$ with $dT=0.1$); 
    hatched region denotes the constant stimulus limit ($d \rightarrow 1$). 
    The $\rho=0.5$ contours correspond to $h(A)=\frac{\alpha}{\beta}\left(\frac{1-q}{1-q^d}\right) q^{d-1}$ where $q=e^{-\alpha T}$.
    }
    \label{fig:H5}
\end{figure}

Although implementing certain hallmarks
turns out to be relatively straightforward, amplitude sensitivity -- which states that weaker stimuli lead to more pronounced habituation -- is more subtle. 
We find that the simplest model proposed above does not satisfy it, 
but a mild extension can (Fig. \ref{fig:H5}). 

It is sufficient to prepend an ``inverter" (or amplitude gate) $u' = h(u)$ that acts on the input signal before it is fed into the linear dynamics for $x$. 
We implement this as 
\begin{equation}
    \label{eq:hallmark5_system}
    \begin{array} {rcl} 
                   u' & = & h(u)\\
               d x/dt & = & \beta u' - \alpha x\\
                    y & = & \textrm{tanh}(\gamma u) \sigma(x) \\
    \end{array}
\end{equation}
\\
where $\sigma(x) = 1/(1 + x^N)$ and $\textrm{tanh}(\gamma u)$ with $\gamma \gg 1$ serves as a sign function so that the response is zero whenever $u=0$, but is passed in a normalized and attenuated manner $y\approx \sigma(x) \in [0,1]$ otherwise.
Note that the static nonlinearities can be viewed as a reduction of a driven nonlinear system $\dot x = f(x,u)$ in $\mathbb{R}^3$, 
as mentioned below Eq. (\ref{eq:filter_A_ode}).

Intuitively, the static input nonlinearity converts strong signals to weak ones and weak signals to strong ones (within some finite envelope). 
Such signal pre-processing may be implemented physically in a variety of ways. For instance, this can be biochemically implemented through rational functions corresponding to enzymatic kinetics, as in Fig. \ref{fig:H5}B. 

Finally, we remark that in the preceding sections, we showed that Eq. (\ref{eq:filter_A_ode}), which involves linear dynamics combined with a static nonlinear output,
exhibits habituation, recovery, and frequency sensitivity (Fig. \ref{fig:H5}D). Here we have shown that generalizing one step further by adding a static nonlinearity to the input as in Eq. (\ref{eq:hallmark5_system}) (i.e. a Hammerstein-Wiener model \cite{Schoukens2017}) allows the system to additionally account for Hallmark 5: amplitude sensitivity (Fig. \ref{fig:H5}C,E).

In summary, the combination of relatively simple dynamical variables combined with static nonlinearities acting on the output (and potentially also the input) is able to account for many of the hallmarks of habituation (Table \ref{tab:hallmarks-main}). 

\vspace{-0.5cm}
\section{Discussion}
Habituation is a fundamental form of nonassociative learning in which a system develops a short-term memory of repetitive stimuli that allows it to attenuate its response to subsequent stimulation. 
This capacity is universally displayed by living systems, where it is thought to facilitate behavioral focus (i.e.\ learning to ignore benign or neutral events). 
While previous studies of habituation have largely focused on neural substrates, observations of habituation in unicellular organisms and even non-living matter demonstrate that neural circuits, while useful, are not required. 
To develop a generalized mathematical framework for describing, detecting, and modeling this phenomenon in seemingly disparate contexts, we have formalized several of the classic hallmarks of habituation in the language of dynamical systems and proposed minimal motifs that can satisfy specific hallmarks. Accordingly, this work establishes a foundation for further theoretical and applied analysis of habituation from a dynamical systems perspective. 

Starting from the simplest possible dynamics, we first note that linear time-invariant systems cannot exhibit both habituation and recovery (which is consistent with a recent study of \emph{Drosophila} larvae habituating to mechanical stimuli \cite{elife2023miami}). 
We then describe how a basic extension of linear dynamics -- introducing a static nonlinearity at the output -- can explain the core hallmarks of habituation and recovery and other more detailed features. 
We further demonstrate that incorporating a nonlinearity also at the input allows the model to additionally account for amplitude sensitivity.
This work establishes minimal motifs for phenomena described thus far largely verbally, despite using linear memory dynamics which are maximally tractable. 

This general structure is compelling because of its simplicity and flexibility, which suggest it could be implemented across scales and experimental contexts. 
For instance, in the setting of unicellular habituation, such motifs can be realized through a generic circuit governing the concentrations of two molecular species ($x$, $y$): 
linear dynamics for the degradation and input-driven synthesis of $x$ serves as a leaky memory of recent stimulation, which can then be used to either (i) slow the production rate or (ii) increase the degradation rate of the response variable $y$. 
In neural settings, series \cite{Staddon1996} and parallel \cite{kohonen1989text} connections of analogous elementary units may be assembled to implement more elaborate information processing and memory (such as frequency sensitivity).

The motifs considered here can readily be mapped to other physical contexts, as shown in \cite{Monnigmann2024CDC} where habituating voltage responses are generated by an RC circuit combined with a resistor-diode loop (acting as a static nonlinear element). 
Furthermore, hallmarks of habituation have also been reported in the electrical conductivity of certain materials, such as various nickel oxides when subjected to repeated hydrogen gas exposure \cite{Zuo2017, Zhang2021resistance}. 
Thus, ongoing efforts to emulate aspects of biological learning via neuromorphic computing \cite{Ramanathan2024} using \emph{physical} habituation substrates are likely to benefit from the presented mathematical framework. 

The identification of minimal mathematical models for conserved biological phenomena has a long history, to which this work contributes for the case of habituation. 
The approach taken here draws inspiration from work on the related phenomenon of adaptation, which has largely been studied from a biochemical perspective. 
As with adaptation, mathematically defining the qualitative property of interest is an important preliminary step toward identifying underlying mechanisms (expressed as dynamical systems) whether by analytic construction, numeric screens \cite{Ma2009}, or evolutionary approaches \cite{Francois2008}. 
This has led to the recognition of numerous motifs for adaptation, such as negative feedback and incoherent feedforward loops, as well as antithetical integral control and other schemes \cite{Tyson2003-SniffersBuzzersEtc, Ferrell201662, Briat201615, Khammash2019}.
Along the same lines, we anticipate that the foundation established here will stimulate the recognition of habituating motifs both mathematically and experimentally in diverse contexts.

\section*{\label{sec:methods}Materials and Methods}

\subsection*{Numerics and Code}
Unless otherwise specified, the numerical trajectories are obtained in Python 3.9.6 using either direct convolution or an implicit Runge-Kutta scheme via the SciPy library. 
The code underlying this work is available at \url{https://github.com/mattsmart/habituation-pub}.

\subsection*{Mathematical details}

\begin{table*}[ht]
    \noindent
    \caption{Hallmarks of habituation from~\cite{Thompson1966, rankin2009} in shortened form, alongside proposed mathematical criteria. 
    Hallmarks after $H_1$ assume there exists a stimulus $u(t)$ for which $H_1$ holds. 
    }

    \begin{tabularx}{\linewidth}{l X X}
        
        \hline
        $H_1$
        & 
        \footnotesize
        \hspace{0.15cm}
        \begin{minipage}[t]{0.95\linewidth}
        \cellcolor{shadecolor}
          \justifying\noindent
          \textbf{Habituation}: 
          Repeated application of a stimulus results in a progressive decrease of a response to an asymptotic level. 
          
        \end{minipage}
        \hspace{0.1cm}
        & 
        \footnotesize
        There exists a periodic stimulus $u(t)$ that generates a sequence of responses satisfying 
        $0 \le y[k+1] \le y[k]$ for all $k\ge 0$, and there exists a $K>0$ such that $y[k+1] < y[k]$ for all $k = 0, \ldots, K$. 
        \\

        \hline
        $H_2$
        & 
        \footnotesize
        \hspace{0.15cm}
        \begin{minipage}[t]{0.95\linewidth}
        \cellcolor{shadecolor}
          \justifying\noindent
          \textbf{Spontaneous recovery}:
          If the stimulus is withheld, the response recovers over time.
        \end{minipage}
        \hspace{0.1cm}
        & 
        \footnotesize
        Assume the stimulus has been applied $k$ times, resulting in a response $y[k]$. 
        There exists an $m\in\mathbb{N}$ such that the response after withholding the stimulus for $m$ periods satisfies $y[k+m+1] > y[k]$.
        \\

        \hline
        $H_3$
        & 
        \footnotesize
        \hspace{0.15cm}
        \begin{minipage}[t]{0.95\linewidth}
        \cellcolor{shadecolor}
          \justifying\noindent
          \textbf{Potentiation of habituation}:
          After multiple series of stimulus repetitions and spontaneous recoveries, the response decrement becomes
          successively more rapid and/or 
          more pronounced. 
        \end{minipage}
        \hspace{0.1cm}
        & 
        \footnotesize
        Assume the stimulus is applied for $L$ periods, subsequently withheld $L^\prime$ periods, and assume this pulse-then-rest pattern is repeated. There exists a number $K\in\mathbb{N}$ 
        of pulse-then-rest repetitions 
        such that 
        $y[K(L+L^\prime)+k]< y[k]$ for some subsequent stimuli  $k= 0, \dots, m-1$, $m>0$. 
        \\

        \hline
        $H_4$
        & 
        \footnotesize
        \hspace{0.15cm}
        \begin{minipage}[t]{0.95\linewidth}
        \cellcolor{shadecolor}
          \justifying\noindent
          \textbf{Frequency sensitivity}:
          Other things being equal, more frequent stimulation results in 
          (a) more rapid and/or more pronounced response decrement, and 
          (b) more rapid recovery.
        \end{minipage}
        \hspace{0.1cm}
        & 
        \footnotesize
        Define $U_T = \{ \textrm{stimulus period\,\,} T\in \mathbb{R}^+ |\, \textrm{$H_1$ holds} \}$.
        If $T_1 < T_2 \in U_T$, then 
        (a) the responses satisfy $y_{1}[k] \le y_{2}[k] \ \forall \ k\in \mathbb{N}$, 
        and 
        (b) $m_1 T_1 \le m_2 T_2$ where $m_i T_i$ defines the time to recover following $k$ stimulations
        (i.e. smallest $m_i$ that satisfies $|y_i[0] - y_i[k + m_i]| < \epsilon$). 
        \\

        \hline
        $H_5$
        & 
        \footnotesize
        \hspace{0.15cm}
        \begin{minipage}[t]{0.95\linewidth}
        \cellcolor{shadecolor}
          \justifying\noindent
          \textbf{Intensity (amplitude) sensitivity}:
          Within a stimulus modality, less intense stimuli give more rapid and/or more pronounced response decrement. 
          Intense stimuli may yield no significant observable response decrement.
        \end{minipage}
        \hspace{0.1cm}
        & 
        \footnotesize
        Define $U_A = \{ \textrm{stimulus intensity\,\,} A\in \mathbb{R}^+ \,|\, \textrm{$H_1$ holds} \}$.
        If $A_1 < A_2 \in U_A$,
        then the responses satisfy $y_{1}[k] \le y_{2}[k] \ \forall \ k\in \mathbb{N}$.
        $U_A$ may be bounded above.
        \\

        \hline
        $H_6$
        & 
        \footnotesize
        \hspace{0.15cm}
        \begin{minipage}[t]{0.95\linewidth}
        \cellcolor{shadecolor}
          \justifying\noindent
          \textbf{Subliminal accumulation}:
          The effects of repeated stimulation may continue to accumulate even after the response has reached an asymptotic level. 
          Among other effects, this can delay the onset of spontaneous recovery.
        \end{minipage}
        \hspace{0.1cm}
        & 
        \footnotesize
        Suppose the response reaches an asymptotic level on the $k^{th}$ stimulation. 
        Let $l_1 < l_2 \in \mathbb{N}$. 
        Then a system receiving $k+l_1$ stimulations will take longer to recover than a system receiving $k+l_2$ stimulations.
        \\

        \hline
        $H_7$
        & 
        \footnotesize
        \hspace{0.15cm}
        \begin{minipage}[t]{0.95\linewidth}
        \cellcolor{shadecolor}
          \justifying\noindent
          \textbf{Stimulus specificity}:
          Within the same stimulus modality, the response decrement shows some stimulus specificity. [...]
        \end{minipage}
        \hspace{0.1cm}
        & 
        \footnotesize
        Not considered. 
        \\
        
        \hline
        $H_8$
        & 
        \footnotesize
        \hspace{0.15cm}
        \begin{minipage}[t]{0.95\linewidth}
        \cellcolor{shadecolor}
          \justifying\noindent
          \textbf{Dishabituation}:
          Presentation of another (usually strong) stimulus results in recovery of the habituated response.
        \end{minipage}
        \hspace{0.1cm}
        & 
        \footnotesize
        There exists a second stimulus $s(t)$ applied only during $kT < t < (k+1)T$ which guarantees $y[k + 1] > y[k]$.
        \\

        \hline
        $H_9$
        & 
        \footnotesize
        \hspace{0.15cm}
        \begin{minipage}[t]{0.95\linewidth}
        \cellcolor{shadecolor}
          \justifying\noindent
          \textbf{Habituation of dishabituation}: Upon repeated application of the dishabituating stimulus, the amount of dishabituation produced decreases.
        \end{minipage}
        \hspace{0.1cm}
        & 
        \footnotesize
        Not considered. 
        \\

        \hline
        $H_{10}$
        & 
        \footnotesize
        \hspace{0.15cm}
        \begin{minipage}[t]{0.95\linewidth}
        \cellcolor{shadecolor}
          \justifying\noindent
          \textbf{Long-term habituation}:
          Some stimulus protocols may result in properties of the response decrement that last hours, days, or weeks.
        \end{minipage}
        \hspace{0.1cm}
        & 
        \footnotesize
        Not considered.
        \\
        
        \hline
        
    \end{tabularx}
    \label{tab:hallmarks-main}
\end{table*}

\subsubsection*{Response to comb of rectangle pulses}
Suppose $u(t)$ is an infinite comb of rectangle pulses with duty $d$, period $T$, and area $\Lambda \equiv A dT$ (amplitude $A=\frac{\Lambda}{dT}$). Assume the first pulse begins at $t=(1-d)T$ and ends at $t=T$. Observe that the limit $d \rightarrow$ 0, $\Lambda=1$ gives a Dirac comb.  

For such inputs, the linear response Eq. (2) (for $x_0=0$ and parameters $\alpha, \beta$) can be characterized by the sequence of minima $\{x_n^{l} \}$ and maxima $\{x_n^{h} \}$ associated to each pulse $n=1,2, \ldots$. 
One obtains the recurrence relations for successive minima and maxima:
\begin{equation}
\label{eq:rect_comb_high_low_recurrence_1}
\begin{array} {rcl} 
             x_n^h & = & x_n^l e^{-\alpha d T} + Q \\
             x_n^l & = & x_{n-1}^h e^{-\alpha T} e^{\alpha d T} \\
\end{array}
\end{equation}
where $Q\equiv \Lambda \beta \frac{1-e^{-\alpha d T}}{\alpha d T}$ is the constant gain from each pulse ($Q \rightarrow \Lambda \beta$ as $d \rightarrow 0$). 
These can be combined to obtain 
\begin{equation}
\label{eq:rect_comb_high_low_recurrence_2}
\begin{array} {rcl} 
             x_n^h & = & x_{n-1}^h e^{-\alpha T} + Q.
\end{array}
\end{equation}

\noindent
Using the initial condition $x_1^l=0$ and substituting $q = e^{-\alpha T} < 1$, we have 
\begin{equation}
\label{eq:rect_comb_high_low_recurrence_3}
\begin{array} {rcl} 
             x_n^h & = & Q \frac{1-q^n}{1-q}.
\end{array}
\end{equation}

\noindent
Thus, the limit cycle for $n\rightarrow \infty$ is characterized by
\begin{equation}
\label{eq:rect_comb_high_low_limitcycle}
\begin{array} {rcl} 
             x_{\infty}^h & = & \frac{Q}{1-q} \\
             x_{\infty}^l & = & x_{\infty}^h q^{1-d}.
\end{array}
\end{equation}

Convergence to the limit cycle is defined using the threshold 
$\left\vert x_{n+1}^h / x_n^h - 1 \right\vert < \epsilon$. 
For fixed $\epsilon$, convergence occurs at the smallest positive integer $n$ such that  $q^n < \frac{\epsilon}{1+\epsilon - q}$,
which gives $n^* = \lceil \ln( \frac{\epsilon}{1+\epsilon - q}) / \ln q \rceil$.

\subsubsection*{Mapping to the discrete-time model of Staddon and Higa}

Refs. \cite{Staddon1993, Staddon1996} propose models of habituation that can be viewed as discrete-time analogs of Eq. (\ref{eq:filter_A_ode}).
These models are based on scalar units of the form
\begin{equation}\label{eq:staddon1993}
\begin{array} {rrcl} 
    \textrm{dynamics}&  x_{t+1} & \!\!\!=\!\!\!\! & a x_t + b u_t\\
    \textrm{output}  &    y_t & \!\!\!=\!\!\!\! & \textrm{ReLU}(u_t - x_t - \theta)\\
\end{array}
\end{equation}
where $0<a<1$ and $b>0$, and a third parameter $\theta$ serves to adjust the response threshold. 
This single-unit model exhibits habituation and recovery, and additionally exhibits some frequency sensitivity when multiple units are chained in series \cite{Staddon1996}. 
As in  Eq. (\ref{eq:filter_A_ode}), $x_t$ is an internal state, $u_t$ is an input time series, and $y_t$ is the output. 
Under mild assumptions on $u(t)$, the dynamics in Eq. (\ref{eq:filter_A_ode}) map directly onto Eq. (\ref{eq:staddon1993}) using a timestep $\Delta t$ and $a\equiv e^{-\alpha \Delta t}$, $b\equiv \frac{\beta}{\alpha} (1 - e^{-\alpha \Delta t})$.

To map from the internal state $x(t)$ of Eq. (\ref{eq:filter_A_ode}) to Eq. (\ref{eq:staddon1993}), consider the difference $x(t+\Delta t) - x(t)$ using the solution of Eq. (\ref{eq:filter_A_ode}) and a fixed timestep $\Delta t$. 
This gives 
\begin{equation}
\begin{array} {rcl} 
    x(t+\Delta t) & = & e^{-\alpha \Delta t} x(t) + \beta e^{-\alpha \Delta t} 
       \int_t^{t+\Delta t} u(\tau) e^{-\alpha (t-\tau)} d\tau \\
                  & = & e^{-\alpha \Delta t} x(t) + \frac{\beta}{\alpha} (1 - e^{-\alpha \Delta t}) u(t) \\
                  & = & a x(t) + b u(t)
\label{eq:map_to_staddon}
\end{array}
\end{equation}
where the second line assumes that $u(t)$ is constant on any interval $t \in [k \Delta t, (k+1) \Delta t]$ and establishes the mapping noted above. 
The forward mapping from the continuous to the discrete dynamics can be generalized to higher dimensions (see e.g. \cite[Sec.~2.5.3]{Hinrichsen2005}), although the reverse mapping may no longer be uniquely defined.

\begin{acknowledgments}
We thank Anton Persikov, Hayden Nunley, Gilles Francfort, Annette Schenck, Marina Boon, Nikolay Kukushkin, and Jasmin Imran Alsous for helpful discussions and suggestions, and Lucy Reading-Ikkanda/Simons Foundation for illustrations.
M.M.\ gratefully acknowledges travel support by the Simons Foundation. M.S.\ and S.Y.S.\ are grateful for ongoing support through the Flatiron Institute, a division of the Simons Foundation. 
\end{acknowledgments}

\begin{center}
    \textbf{Author contributions}
\end{center}
  Author contributions:
    conceptualization, M.S., S.Y.S., M.M.; 
    methodology, M.S., M.M.; 
    investigation, M.S., M.M.; 
    writing – original draft, M.S., M.M.; 
    writing – review \& editing, M.S., S.Y.S., M.M.

\begin{center}
    \textbf{Declaration of interests}
\end{center}
The authors declare no competing interests.

%
. 

\end{document}